\documentclass[a4paper,11pt]{article}
\pdfoutput=1
\usepackage{jheppub}
\usepackage[T1]{fontenc}
\usepackage{dcolumn}% Align table columns on decimal point
\usepackage{bm}% bold math
\usepackage[utf8]{inputenc}
\usepackage{graphicx}
\usepackage{hyperref}

\usepackage{xcolor}

\def\nn{\nonumber}

\title{Searching for Charged Higgs Bosons 
via $e^+ e^- \to H^+ H^- \to c\bar{b} \bar{c}b $ at Linear Colliders
}

\author[a]{Wei-Shu Hou}
\author[a]{Rishabh Jain}
\author[b]{Tanmoy Modak}

\affiliation[a]{Department of Physics, National Taiwan University, Taipei 10617, Taiwan}
\affiliation[b]{
Institut f\" ur Theoretische Physik, Universit\"at Heidelberg, 69120 Heidelberg, Germany}
\emailAdd{wshou@phys.ntu.edu.tw}
\emailAdd{rishabhjain@hep1.phys.ntu.edu.tw}
\emailAdd{tanmoyy@thphys.uni-heidelberg.de}

\abstract{
We study a search for the charged Higgs boson via
 $e^+e^- \to H^+H^- \to c\bar{b}\bar{c}b$ at the 500\,GeV\;ILC. 
In a general two Higgs doublet model without $Z_2$ symmetry, 
extra Yukawa couplings $\rho_{tt}$ and $\rho_{tc}$ can drive baryogenesis, 
but searches at the HL-LHC may still go empty-handed if the couplings are relatively weak. 
Taking $m_{H^+ } \simeq m_H \simeq m_A \simeq 200$\;GeV, 
with $\rho_{tt}$, $\rho_{tc}\sim 0.1$ and no $h(125)$-$H$ mixing, 
$H^+ \to c\bar b$ decay is dominant, and the $c\bar{b}\bar{c}b$ final state 
is likely overwhelmed by QCD background at the LHC.
We show that the electroweak production of $H^+ H^-$ at the ILC 
is discoverable with integrated luminosity of 1\,ab$^{-1}$.
Furthermore, we show that $m_{H^+}$ can be extracted by requiring 
the two pairs of $b$ and light jets be roughly equal in mass, 
without assuming the mass value. 
Thus, ILC can probe low mass Higgs bosons in multijet final states 
to complement HL-LHC in the future.}
\arxivnumber{2111.06523}

\begin{document}

\maketitle

\flushbottom

\section{Introduction}

%-----------------------------------------------------------------------------------------
% Section I
%----------------------------------------------------------------------------------------

Many extensions of the Standard Model (SM) predict 
the existence of extra Higgs bosons, for example,
%, which can help our  understanding of electroweak symmetry breaking (EWSB). 
%These extra Higgs bosons can
to provide additional $CP$ violation required for baryogenesis, 
to account for matter dominance of the Universe~\cite{Fuyuto:2017ewj,Chen:2017com}. 
In this report we consider the general two Higgs doublet model 
(g2HDM) without $Z_2$ symmetry, where after 
electroweak symmetry breaking (EWSB) one has 5 Higgs bosons:
 two $CP$-even neutral scalars $h, H$, one $CP$-odd scalar $A$, 
 and charged scalars $H^{\pm}$. 
Without a $Z_2$ symmetry, one has an extra set of Yukawa couplings for each type of charged fermion $f$. In general, 
{\boldmath $\rho$}$^f$ cannot be diagonalized simultaneously with 
the mass matrix {\boldmath $m$}$^f$, 
leading to flavor changing neutral couplings (FCNC). 
As these are usually considered as dangerous, Natural Flavor Conservation 
was proposed~\cite{NFC_glashow} to kill the extra Yukawa couplings,
hence avoid FCNCs altogether. 
We advocate, however, that one should leave 
the final say on the matter to the experiments. 
 
It was shown~\cite{Fuyuto:2017ewj, Fuyuto:2019svr} that 
extra top-Yukawa couplings $\rho_{tt}$ and $\rho_{tc}$ 
can drive electroweak baryogenesis (EWBG). 
Following this motivation, it was pointed out that 
$cg \to tH/tA \to tt\bar c, tt\bar t$~\cite{Kohda:2017fkn} 
can provide same-sign top and triple-top signatures at the LHC. 
However, $\mathcal{O}(1)$ values of $\rho_{tc}$ is 
constrained by~\cite{Hou:2018zmg,Hou:2019gpn} from CRW, 
a Control Region for $t\bar tW$ background in 
the 4-top study by CMS~\cite{CMS:2017ocm}. 
On the other hand, $\rho_{tt}$ is constrained by 
$b\to s\gamma$ and $B$--$\bar{B}$ mixing~\cite{Altunkaynak:2015twa}. 
All these limits are modulated by the unknown scalar masses, 
except the already observed $h(125)$ \cite{ATLAS:2012yve,CMS:2012qbp}. 
With rich dynamics and sub-TeV extra Higgs bosons~\cite{Hou:2017hiw}, 
the g2HDM can provide very rich signatures at the LHC~\cite{Hou:2020chc} 
and impact on flavor physics~\cite{Hou:2020itz}.

ATLAS and CMS have performed many searches 
for extra scalars in different channels, such as
 $H/A \to \tau\tau$~\cite{CMS:2018rmh,ATLAS:2020zms},
 $H \to \tau \mu$~\cite{CMS:2019pex},
 $H/A \to b \bar{b}$~\cite{CMS:2018hir,ATLAS:2019tpq},
 $H,A\to t\bar{t}$~\cite{ATLAS:2017snw,CMS:2019pzc},
 $H^+ \to t\bar{b}$~\cite{CMS:2020imj,ATLAS:2021upq},
 $H^+ \to cs$~\cite{ATLAS:2013uxj,CMS:2020osd},
 $H^+ \to \tau\nu$~\cite{ATLAS:2018gfm,CMS:2019bfg},
%}
% ~\cite{ATLAS:2017snw,ATLAS:2020zms,ATLAS:2018gfm,
%CMS:2020osd,CMS:2020ffa,CMS:2019bfg,CMS:2018dzl,CMS:2018rkg,
%ATLAS:2021hbr,ATLAS:2021upq}
and over broad mass ranges. 
But a scalar of mass around 200\;GeV decaying into jets 
is relatively challenging for the LHC due to high QCD backgrounds. 
Furthermore, cross sections become challenging at the LHC if the 
extra top Yukawa couplings are considerably below ${\cal O}(1)$~\cite{Hou:2020chc}.
For the special case of sizable $\rho_{tc}$ but all other extra Yukawa couplings negligible in strength, 
the cancellation between $H$ and $A$ for $cg \to tH/tA \to tt\bar c$~\cite{Kohda:2017fkn,Hou:2020chc}
again makes the extra scalars unobservable at the LHC.
In this report, we propose that if a light $H^+$ decaying hadronically
remains elusive at the High Luminosity LHC (HL-LHC),
a linear $e^+ e^-$ collider with sub-TeV energies would be 
able to probe $m_{H^+}$ at ${\cal O}$(200) GeV through 
$e^+ e^- \to H^+ H^- \to c\bar{b}\bar{c}b$. Similar processes have been proposed previously for
$m_{H^+} \lesssim M_Z$~\cite{Akeroyd:1994ga,Akeroyd:2018axd,Akeroyd:2019mvt}.

\begin{figure*}[t]
  \centering
  \includegraphics[scale=0.21]{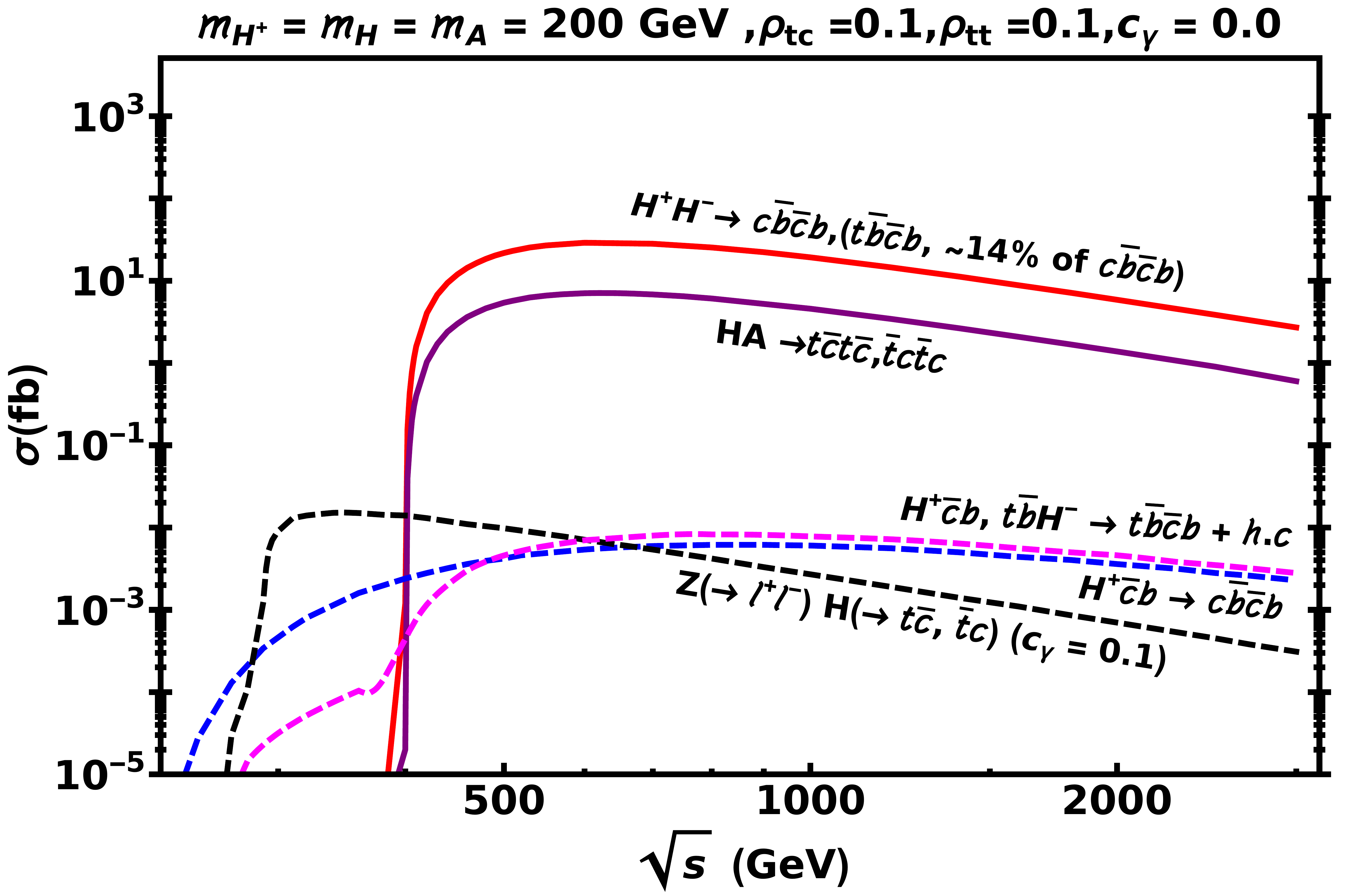}
\caption{
Cross section vs $\sqrt{s}$ for
  $e^+e^- \to H^+ H^- \to c\bar{b}\bar{c}b$ (red, solid),
  $HA \to t\bar{c} t\bar{c}, \bar{t} c\bar{t} c$ (purple solid),
  ${Z(\to \ell\ell)}H(\to t\bar{c}, \bar{t}c$)\; (black, dashed, with $c_\gamma = 0.1$),
  {$H^{+}\bar cb \to c\bar{b} \bar{c}b$} (blue, dashed), and
  {$H^{+}\bar cb,\, t\bar bH^{-} \to t\bar b\bar cb + {\rm h.c.}$} (magenta, dashed). 
 Note here that the $c_\gamma$ is nonvanishing only for ${Z(\to \ell\ell)}H(\to t\bar{c}, \bar{t}c)$ process but 
   set to zero for all other curves and throughout the manuscript.
}
  \label{fig:crossx}
\end{figure*}

We take $\rho_{tc} = \rho_{tt} = 0.1$ to evade HL-LHC 
search~\cite{Hou:2020chc}. 
We assume degenerate $m_{H^+ } = m_H = m_A = 200$\;GeV,
and take the alignment limit of vanishing $h$-$H$ mixing
 ($c_\gamma \equiv \cos\gamma \to 0$)
to reduce the number of parameters.
Our focus will be the electroweak production process
\begin{equation}
 e^+e^- \to \gamma^*/Z^* \to H^+H^- \to c\bar{b}\bar{c}b.
\label{eq:eeH+H-}
\end{equation}
Then $\rho_{tc} = \rho_{tt} \sim 0.1$ means 
$t\bar b\bar cb$ (plus conjugate) cross section is at 
14\% of $c\bar b\bar cb$, as illustrated in Fig.~\ref{fig:crossx}, 
and provides a probe of $\rho_{tt}$.
But for $|\rho_{tt}| \ll |\rho_{tc}|$,
the process of Eq.~(\ref{eq:eeH+H-}) becomes independent of $\rho_{tc}$, 
provided it is the single dominant coupling.
However, the 4-jet final state may appear worrisome for mass extraction,
hence interpretation itself of the origin of these events.

As far as interpretation is concerned, it was suggested~\cite{Hou:1995qh} 
long ago that the companion process
\begin{equation}
 e^+e^- \to Z^* \to HA \to t\bar{c}t\bar{c},\, \bar{t}c\bar{t}c,
\label{eq:eeHA}
\end{equation}
can provide same-sign top plus two jet signature,
with cross section (see Fig.~\ref{fig:crossx}) 
$\sim 1/3$ of $c\bar{b}\bar{c}b$ from $H^+H^-$ production, 
which can support the g2HDM interpretation.
But it would not help mass extraction, as same-sign top
requires both tops to decay semi-leptonically.

%-------------------------------------------
%  Figure 1, Production cross-section diff processes
%-------------------------------------------

For sake of helping the extraction of exotic Higgs mass,
by analogy with $Zh$ production, where $Z \to \ell\ell$ 
would tag the recoil mass, the process
%
%\begin{equation}
 $e^+e^- \to Z^* \to ZH$
%\label{eq:eeZH}
%\end{equation}
%
comes to mind. Unfortunately, this process is suppressed by 
$c_\gamma^2$ (see  Fig.~\ref{fig:crossx} for the case of $c_\gamma = 0.1$),
and statistics would likely be insufficient.

As a final attempt, one has associated production of single $H^+$, 
i.e.
%
%\begin{equation}
 $e^+e^- \to Z^* \to H^+\bar cb \to c\bar b\bar cb, t\bar b\bar cb\, (+\,{\rm h.c.})$ and
 $e^+e^- \to Z^* \to t\bar bH^- \to t\bar b\bar cb$ + h.c.,
where the latter has a higher threshold.
%\label{eq:bH+ct} 
%\end{equation}
%
The cross sections are also given in Fig.~\ref{fig:crossx}, 
which turn out tiny because of the three-body nature and
destructive interference between $Zbb$ and $Zcc$ diagrams. 
Being only a small fraction of the main process, Eq.~(\ref{eq:eeH+H-}),
these can be safely ignored.

Although not intuitive at first sight, it turns out
the 4-jet process of Eq.~(\ref{eq:eeH+H-}) 
does allow one to extract $m_{H^+}$.
With two jets $b$-tagged {and the $c$-jets untagged (achievable at ILC),} 
one has two possible $bj$ pairings. 
By demanding the two $bj$ pairs to be close in mass,
{\it without} specifying the mass, 
one not only rejects background effectively, 
the underlying two-body production kinematics 
allow the extraction of $m_{H^+}$, as we will show.

This report is organized as follows. 
We first give the framework of our study in Sec.~II.
We describe our signal from Eq.~(\ref{eq:eeH+H-}) 
and all background processes in Sec.~III, 
and state our event selection criteria and analysis strategy. 
In Sec.~IV, we discuss the discovery potential, 
method of mass extraction, 
and the mass window cut. 
After some discussions, we conclude in Sec.~V.

\section{Framework and Signal Process}

For CP conserving Higgs sector, one can write the Higgs potential of g2HDM 
the Higgs basis as~\cite{Hou:2017hiw,Davidson:2005cw}
\begin{eqnarray}
  V(\Phi,\Phi^{'}) = \mu_{11}^2 |\Phi|^2 + \mu_{22}^2 |\Phi^{'}|^{2} - (\mu_{12}^2 \Phi^{\dagger} \Phi^{'} + h.c) \nn 
  + \frac{1}{2}\eta_1 |\Phi|^4 + \frac{1}{2}\eta_2 |\Phi^{'}|^4 \\ +   \eta_3 |\Phi|^2|\Phi^{'}|^2 + \eta_4 |\Phi^{\dagger}\Phi^{'}|^2   
  +  \left[ \frac{1}{2}\eta_5 (\Phi^{\dagger}\Phi^{'})^2 + (\eta_6 |\Phi|^2 + \eta_7 |\Phi^{'}|^2)\Phi^{\dagger}\Phi^{'} + h.c.\right],
\label{eq:Pot}
\end{eqnarray}
where EWSB arises from $\Phi$ while $\langle\Phi'\rangle = 0$ (hence $\mu_{22}^2 > 0$).
In Eq.~(\ref{eq:Pot}), $\eta_i$s are the quartic couplings and taken as real,
as we assume the Higgs potential is $CP$-invariant. 
After EWSB, one can find~\cite{Hou:2017hiw} from Eq.~(\ref{eq:Pot})
the mass eigenstates $h$, $H$, $A$ and $H^+$,
%$m_h^2 = \eta_1 v^2$, $m^2_{H^{\pm}} = \mu_{22}^2 + \frac{1}{2} \eta_3 v^2$, and %$m_{H,A} = m^2_{H^{+}} + \frac{1}{2} (\eta_4 \pm \eta_5)v^2$. 
% 
as well as $h$-$H$ mixing, where we define the mixing angle as $\gamma$.

The Yukawa couplings of the Higgs bosons to fermions
are given as~\cite{Davidson:2005cw,Chen:2013qta},
\begin{eqnarray}
 -\frac{1}{\sqrt{2}}\sum_{f = u,d} & \bar{f}_i
  \left[(\lambda_{i}^f\delta_{ij}c_{\gamma} - \rho^{f}_{ij}s_{\gamma})H
         - i\,\mathrm{sgn}(\mathcal{Q}_f)\rho^{f}_{ij}A \right] R f_{j} \nn \\
 -& \bar{u}_{i}\bigl[(V\rho^d)_{ij} R - (\rho^{u\dagger}V)_{ij} L\bigr]d_{j}H^{+}
 + h.c.,
\label{eq:Yuk}
\end{eqnarray}
where $i, j$ are summed over, $\lambda_i^f = \sqrt{2}\, m^f_i/v$ 
is the Yukawa coupling in SM, and 
$c_\gamma\, (s_\gamma) \equiv \cos\gamma\, (\sin\gamma)$. 
From Eq.~(\ref{eq:Yuk}) one finds that $\bar cbH^+$ and $\bar tbH^+$ 
couple as $\rho_{tc}V_{tb}$ and $\rho_{tc}V_{tb}$, 
i.e. {\it both} with CKM factor $V_{tb}$,
where we have dropped CKM-suppressed terms.
This may seem counterintuitive from the perspective of 
say 2HDM~II~\cite{Branco:2011iw},
the popular 2HDM with $Z_2$ symmetry that automatically arises with supersymmetry,
where one expects the $\bar cbH^+$ coupling to be suppressed by $V_{cb}$.
It was through these couplings that $cg \to bH^+ \to bt\bar b$ was 
proposed~\cite{Ghosh:2019exx} as a search mode for $H^+$ production,
and traces back to similar arguments~\cite{Hou:2019uxa} for 
the $V_{tb}/V_{ub}$ enhancement of 
the $\rho_{tu}$ coupling in $B^+ \to \mu^+\nu$ decay.

Experimental searches suggest that $h(125)$ closely resembles the Higgs boson of SM, 
i.e. approximate alignment~\cite{CMS:2018uag,ATLAS:2019nkf}. 
This does not~\cite{Hou:2017hiw} necessarily imply 
that the sole mixing parameter, $\eta_6$, has to be small.
However, as we anticipate the case of low mass and near-degeneracy
(in part to reduce the number of parameters),
\begin{align}
m_{H^+} \simeq m_H \simeq m_A \simeq 200\,{\rm GeV},
\label{eq:low_degen}
\end{align}
to evade HL-LHC by decay to jets,
alignment needs~\cite{Hou:2017hiw} $\eta_6$ to be small.
We take $c_\gamma = 0$ again for simplicity,
which can be easily put back into experimental analysis.

We can now see the reason of choosing Eq.~(\ref{eq:low_degen}),
together with $\rho_{tc} \sim \rho_{tt} \sim 0.1$, 
as possibly evading HL-LHC scrutiny. The $cg \to tH/tA  \to tt\bar c$ 
would suffer cancellation~\cite{Kohda:2017fkn,Hou:2020chc}
from the degeneracy while suppressed by $|\rho_{tc}|^2 \sim 0.01$; 
the $cg \to bH^+ \to bc\bar b$ process would dominate 
over $bt\bar b$, as $H^+ \to t\bar b$ is kinematically suppressed.
But a $bc\bar b$ final state would clearly be swamped by QCD background.
Noting that the same argument would hold for the case of
finite $\rho_{tc}$ but all other $\rho_{ij}$'s negligible,
we turn to the ILC at 500~GeV.

%We have already discussed the four processes in Fig.~\ref{fig:crossx}.
Our signal process of Eq.~(\ref{eq:eeH+H-}) 
is electroweak production of the $H^+H^-$ pair.
The similar process at LHC with $c\bar b\bar cb$ final state 
is clearly swamped by QCD background.

A nonzero $\rho_{tc}$ can drive the same-sign top channel 
via $H, A \to t\bar c, \bar tc$ in the process of Eq.~(\ref{eq:eeHA}), 
where the {$HA$ is produced via $Z$} boson only, 
with $ZHA$ coupling $\propto s_{\gamma}$~\cite{Hou:1995qh}. 
Although the same-sign top signature would make clear the g2HDM origin,
but as argued, having missing energy and mass due to two 
unseen  neutrinos in the final state, $m_{H^+}$ reconstruction is not possible.

Let us not repeat the discussion already given in the Introduction,
but just note from Fig.~\ref{fig:crossx} that the two other processes 
brought in to help salvage mass reconstruction
both turn out to have too small cross sections to be relevant.
But interestingly, the main process of  Eq.~(\ref{eq:eeH+H-})
turns out to actually allow $m_{H^+}$ extraction, which utilizes 
the kinematics of effective two-body production, as we will show.
But let us first go through the signal of  Eq.~(\ref{eq:eeH+H-}) 
and the corresponding background processes.

%-------------------------------
% Section III
%-------------------------------

\section{Signal and background processes}

For our signal channel of $e^+ e^- \to H^+ H^- \to c\bar{b}\bar{c}b$, 
we consider a final state of two $b$-jets plus two light jets. 

We use Madgraph~\cite{Alwall:2011uj} to generate the samples 
for the signal using the 2HDM model formulated via Feynrules~\cite{Alloul:2013bka}. 
We consider all the interfering contributions from SM with our signal, which 
includes $ZZ, Zh$, and we assume unpolarized beams in our study. 
In addition, we include the single charged Higgs contribution $H^+\bar cb$. 
Nevertheless, as we have discussed, this single $H^+$ associated production 
is fairly suppressed, and we do not attempt to extract it.
We use PYTHIA6.4 for hadronization and modeling of initial state radiation, 
as discussed in Ref.~\cite{Sjostrand:2006za}. 
We then pass the sample to Delphes3.5.0~\cite{deFavereau:2013fsa} 
for fast detector simulation, using the default 
International Linear Detector (ILD) card~\cite{Behnke:2013lya}, 
as well as the anti-$k_T$ algorithm~\cite{Cacciari:2008gp} for jet reconstruction. 

Whether interfering or not, we generate events for 
the following backgrounds:
 $t\bar{t} \to b\tilde{j}\tilde{j} \bar{b}\tilde{j} \tilde{j}$; 
 $W^+W^-\to 4 \tilde{j}$;
 $Zh$, $ZZ\to jjb\bar{b}$.
Here $j$ is a light jet and $\tilde j = j$ or $c$. 
We also include backgrounds from $Z \to b\bar b\; (jj)$ where 
a gluon is radiated and splits into $jj$\;($b\bar b$), which 
we denote as $b\bar{b} jj\, (\mathrm{QCD})$. 
Here, $j$ could also be a gluon; replacing $b$ by $c$, 
we get $c\bar{c} \tilde j\tilde j\, (\mathrm{QCD})$ background, 
where $c$ is mistagged as $b$  where mistag rate depends on the transverse momentum of a $c$ as described in the default ILD card~\cite{Behnke:2013lya}.

{Tagging of $b$ brings out a special category 
that we call $b\bar bb\bar b$, which can be fed by
$Zg^*$, $Zh$ and $ZZ$, where $Zg^*$ is 
the aforementioned QCD process but with $jj = b\bar b$,
where we let $b$-tagging run its course
(a similar effect for $c\bar cc\bar c$ is absorbed 
into $c\bar{c} \tilde j\tilde j\, (\mathrm{QCD})$).}
We follow the same procedure as signal to generate background events. 

We choose $m_{H^+} = m_H = m_A = 200$ GeV and 
set all $\rho_{ij} = 0$, except $\rho_{tc} = \rho_{tt} = 0.1$, with $c_\gamma = 0$. 
These choices force $H^+, A, H$ to decay only via $\rho_{tc}$ and $\rho_{tt}$, 
i.e. $H^+ \to c\bar{b},\, t\bar{b}$ and $H, A \to t \bar{c},\, \bar{t} c$. 
The decay widths are 0.127, 0.008 and 0.008 GeV, respectively. 

To select the events, we require
\begin{itemize}
  \item $p_T(j,b) >$ 20 GeV, $|\eta(j,b)| < 2.5$; 
  \item Number of $b$-jets and light jets: $N_{b}$ = 2, $N_{j}$ = 2; 
  \item Number of isolated leptons = 0.
 \end{itemize}
Note that the $p_T$ and $\eta$ cut are intrinsic cuts of the ILD card. 
The requirement of no isolated leptons removes most of 
the $t\bar{t} \to bjjb\ell\nu$ background (labeled as $tt\, 4\tilde j$). 
We denote the above selection requirements as Cuts A.

%--------------------------------------------
%. Figure 2, m_{bj} following the cut flow
%--------------------------------------------
\begin{figure*}
  \centering
  \includegraphics[scale=0.35]{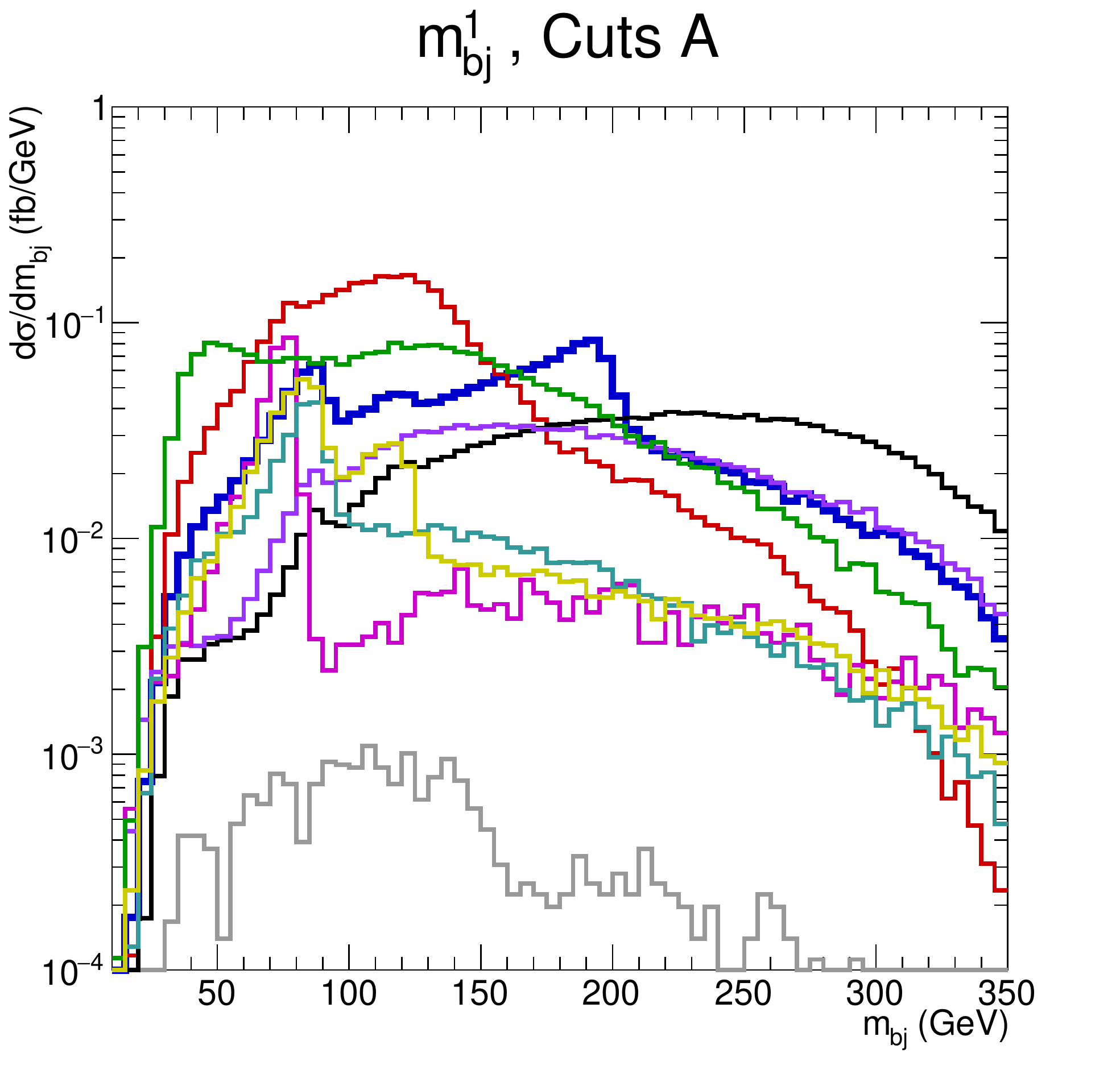}
  \includegraphics[scale=0.35]{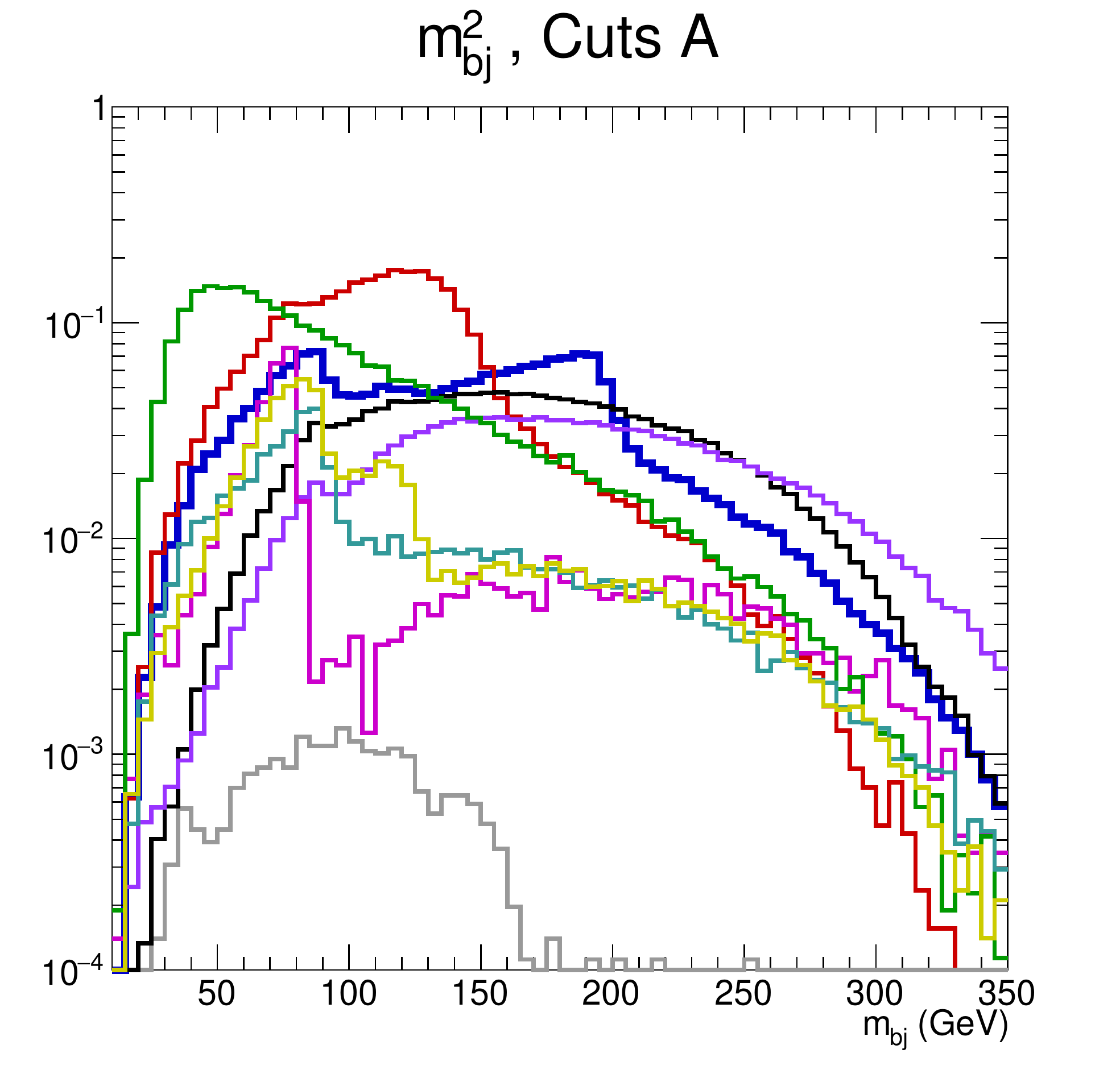} \\
  \includegraphics[scale=0.35]{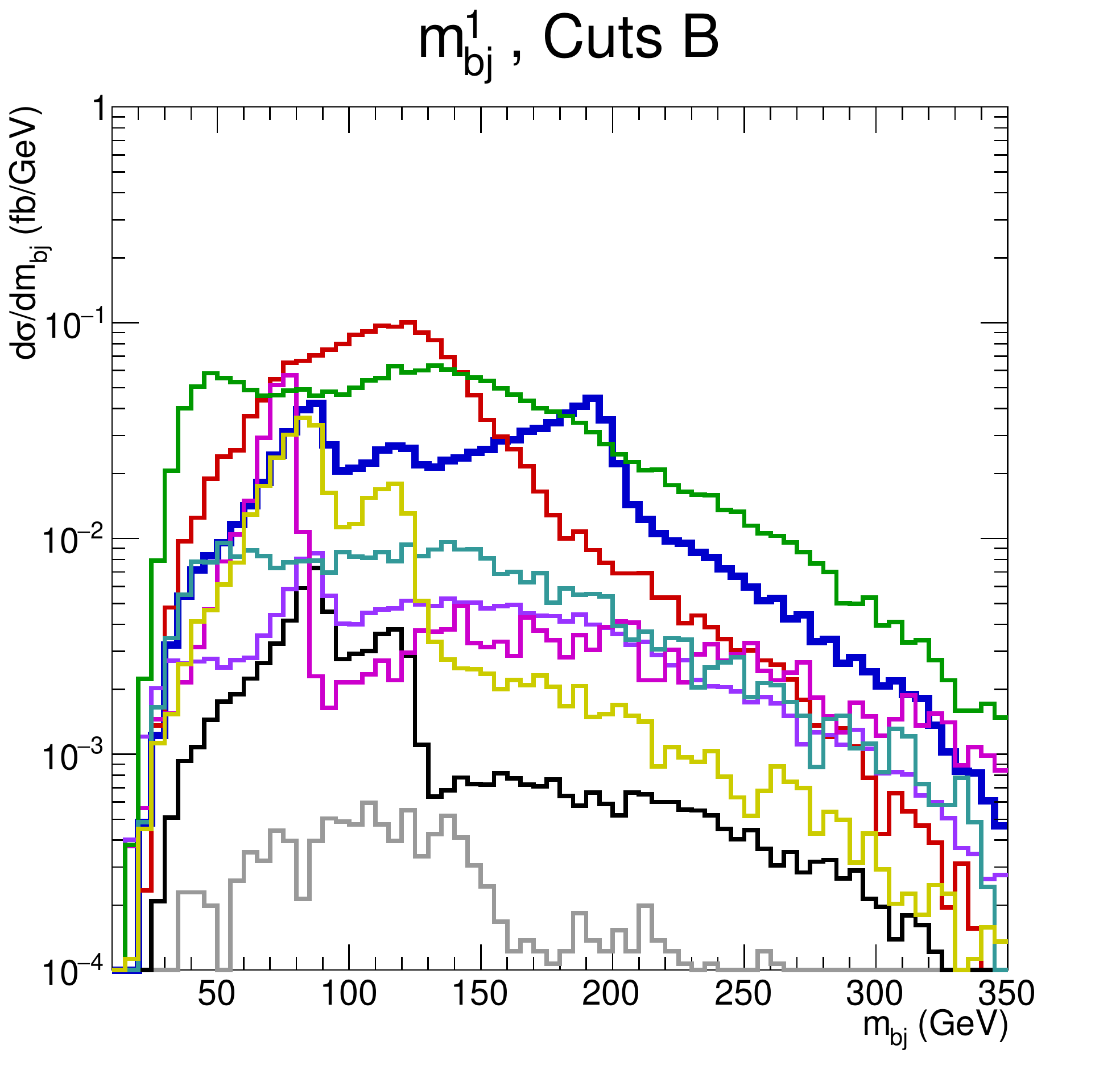}
  \includegraphics[scale=0.35]{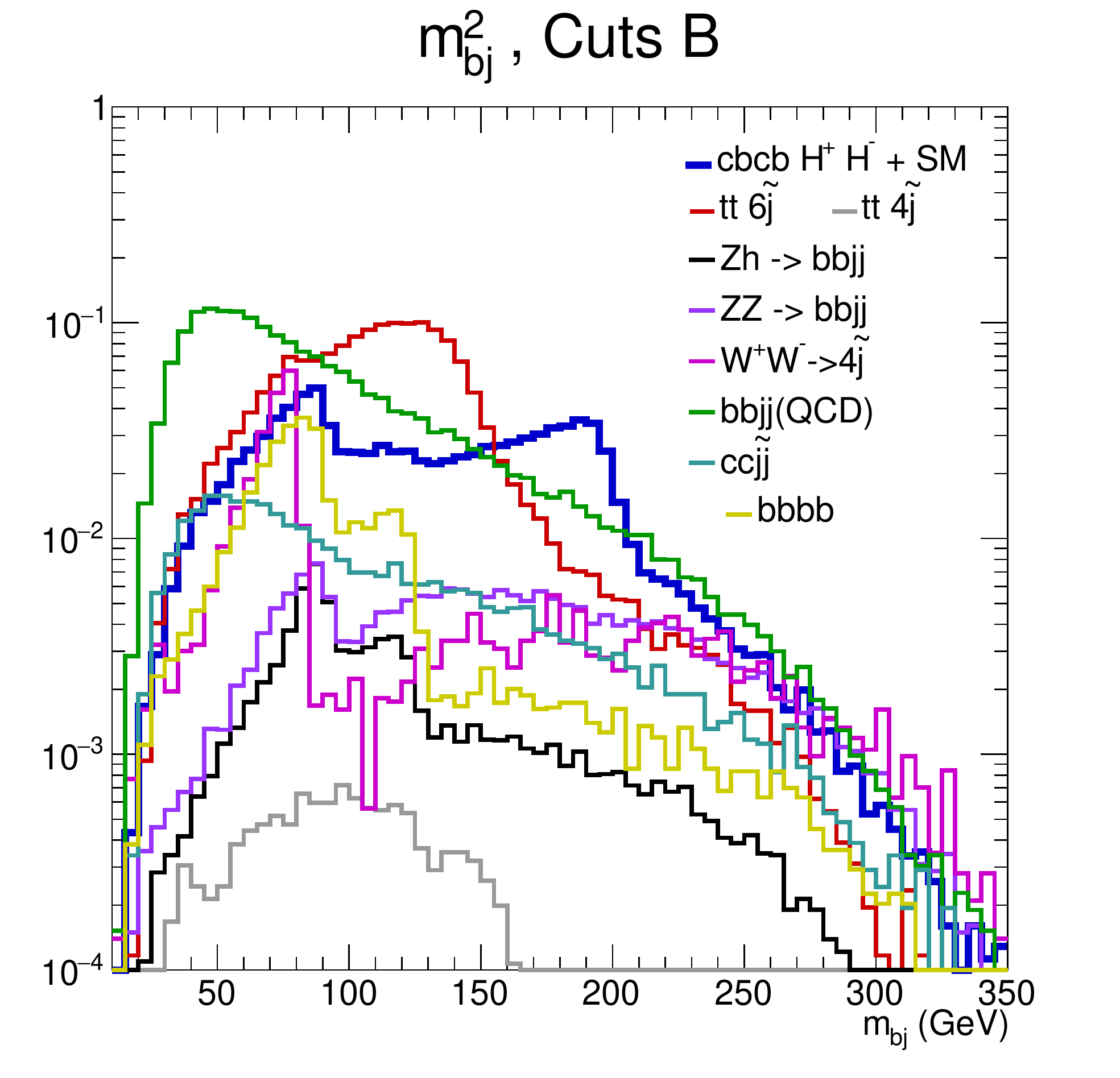}
\caption{
 Reconstructed $m_{bj}$ from both $bj$ pairs, following 
 Cuts A and B. See text for more details.}
 \label{fig:Kin}
\end{figure*}

\begin{table}[b]
  \centering
\begin{tabular}{crr} \hline\hline
  Process & \ Cuts A \ & \ Cuts B \ \\ \hline 
   $c\bar{b}\bar{c}b$ ($H^+H^-$+\,SM) & 10.2 fb \ & 5.0 fb \ \\ 
   $c\bar{b}\bar{c}b$ (SM-only) & \ 4.9 fb \ & 1.7 fb \ \\ 
   \hline
   $t\bar{t}$ & 15.3 fb \ & 8.3 fb \ \\
   $b\bar{b}jj$\,(QCD) & 13.1 fb \ & 9.7 fb \ \\
   $c\bar{c}\tilde{j}\tilde{j}$\,(QCD) & \ 2.7 fb \ & \ 1.6 fb \ \\
   {$b\bar{b}b\bar{b}$} & \ 3.3 fb \ & \ 1.7 fb \ \\
   $ZZ$   & \ 6.3 fb \ & 1.0 fb \ \\
   $Zh$   & \ 7.8 fb \ & 0.4 fb \ \\
   $W^+W^-$   & \ 2.5 fb \ & 1.7 fb \ \\
   % ttH  & &  \\
   \hline
   Total (SM-only) & 55.8 fb \ & 26.1 fb \ \\
   Total ($H^+H^-$+\,SM)  & 61.1 fb \ & 29.5 fb \ \\
  %  $S/B$ & 0.1 & 0.16 \\ 
   Significance ($\mathcal{L} = 1\,{\rm ab}^{-1}$) & 22.1 \,\ \ & 20.2 \,\ \
  \\ \hline\hline
  \end{tabular}
  \caption{
  Cross sections in fb of the interfering Signal and rest of the backgrounds. 
  We also show $c\bar{b}\bar{c}b$ contribution for SM only after Cuts A and Cuts B.
  See text for more details}
  \label{tab:cutAB}
\end{table}

Next, we veto all events that satisfy $|m_{bb} \,(m_{jj}) - m_Z| <$ 15\;GeV, 
$|m_{bb} - m_h| <$ 20\;GeV to reduce $Zh$ and $ZZ$ backgrounds, 
where $m_{bb}\,(m_{jj})$ is the reconstructed mass of 
two $b$-tagged ($j$) jets. We denote this $Z$, $h$ veto as Cuts B. 
In Table~\ref{tab:cutAB} we present our estimates of the cross sections  after applying Cuts~A and B. 
We could apply a cut to veto the events with $m_W$, 
but since $W^+$ and $H^+$ masses are far apart, 
we do not gain much in doing so.

In Fig.~\ref{fig:Kin}, we show in the top and 
bottom frames the reconstructed $m_{bj}$ pairs 
after applying Cuts A and Cuts~B. 
Here we select a $b$-jet and $j$ which are in closer angular proximity to each other.
This separates the two pairs, and gives rise to some mild difference 
between left and right panels.
Our attempt to access the mass of $H^+$ is discussed in the next section.

We also give our significance estimates in Table~\ref{tab:cutAB}.
Since our signal interferes with SM ({at a negligible 2\%}), 
we first estimate the SM only contribution to $c\bar{b}\bar{c}b$ final state 
and combine this with the other non-interfering backgrounds as discussed above. 
We call this total cross section as the SM-only contribution. 
We estimate significance with $\mathcal{L} = 1\, {\rm ab}^{-1}$. 
Suppose we have $n_{\rm pred}$ events from SM, 
while including the $H^+ H^-$ contribution we have $n$ events. 
The significance is then estimated using the likelihood of 
a simple counting experiment as~\cite{Cowan:2010js},
\begin{equation}
    Z(n|n_{\rm pred}) = \sqrt{-2\,ln\frac{L(n|n_{\rm pred})}{L(n|n)}},
\label{eq:Cowan}
\end{equation}
where $L(n_1|n_0) = e^{-n_1}n_1^{n_0}/n_{0}!$.

%---------------------------------------------
%. Section 4 Discovery Potential + Mass rec
%--------------------------------------------
\section{Prospects for Discovery and Mass Extraction}

In this section we discuss the prospect for discovery of our signal. 
We see from Table~\ref{tab:cutAB} that the $H^+H^-$ signal 
can enhance the $bbjj$ final state significantly.
But to demonstrate we do have $H^+ \to c\bar b$,
we attempt at further improvements
with mass extraction in mind.

%--------------------------------------------
%. Figure 3, m_{bj} following Cut C
%--------------------------------------------
\begin{figure*}
  \centering
   \includegraphics[scale=0.35]{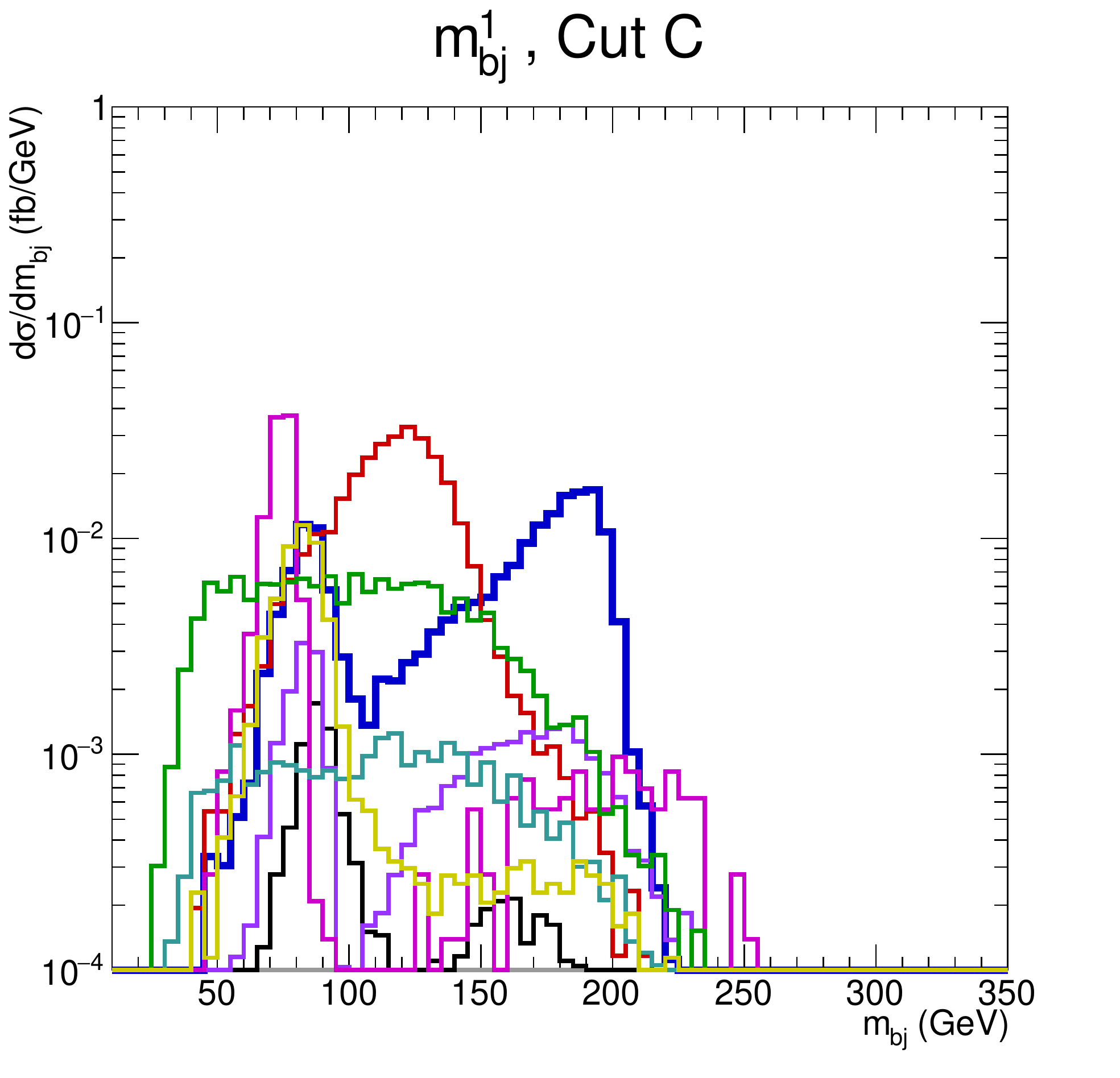} 
   \includegraphics[scale=0.35]{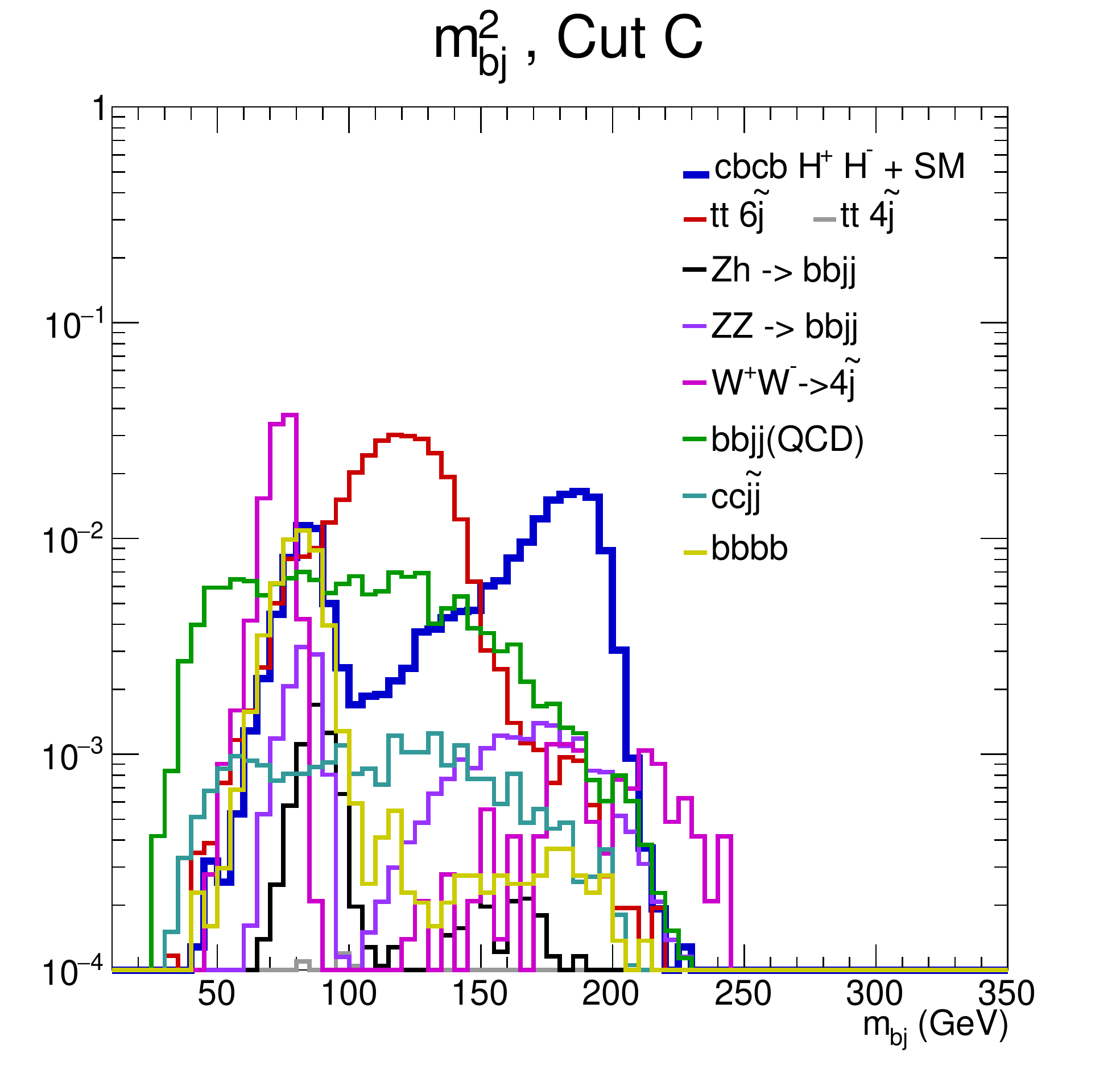}
\caption{
 Reconstructed $m_{bj}$ from both $bj$ pairs, following 
 Cut C, Eq.~(\ref{eq:pairing}). See text for more details.}
 \label{fig:CutC}
\end{figure*}

We first tried the scalar sum, $H_T$, of the strength of $p_T$ of all four jets 
by requiring $H_T \geq $ 300~GeV. %and add this cut after Cuts B. 
This improves our significance from $\simeq 21\sigma$ to 22.5$\sigma$ 
as compared with Cuts B, but it is not so different from Cuts A.
For our second approach, 
we note that to extract the charged Higgs mass, 
one needs to identify the correct bottom and charm pairing, 
to which we turn to discuss.

%-----------------------------------------------
% Mass recovery method
%-------------------------------------------

{\it \textbf{\boldmath Towards\;$m_{H^+}$\,Extraction}.---} \ 
The $H^+H^-$ pair turns into four-jets, 
{and one could misidentify the pairing}. 
To identify the correct pairing, we first select one $b$-tagged jet, 
and chose from the two light jets the one 
at closer angular proximity to the selected $b$-jet. 
This is not very effective for the 500 GeV ILC.
We denote this reconstructed pair mass $m_{bj}^1$ 
and the remaining pair as $m_{bj}^2$, 
which is what is plotted in Fig.~\ref{fig:Kin}. 
Note that Delphes orders the jets by $p_T$, so our choice causes a certain bias 
that makes the two distributions a little different, as already mentioned. 
To improve the right pairing, we then apply the condition 
\begin{equation}
|m_{bj}^{1} - m_{bj}^{2}| < 0.1\times m_{bj}^{1}. 
\label{eq:pairing}
\end{equation}
The 0.1 factor is somewhat arbitrarily chosen for sake of retaining statistics, 
with no attempt at optimization to make $m_{bj}^1 = m_{bj}^2$ more restrictive.
We call the approximate equality condition of Eq.~(\ref{eq:pairing}) Cut C. 
The distribution of $m_{bj}^1$ and $m_{bj}^2$ %
{after applying Cut C} 
is presented in Fig.~\ref{fig:CutC}, with %
{cross sections given in Table~\ref{tab:cutCD}.}

\begin{table}[b]
  \centering
\begin{tabular}{cll} \hline\hline
  Process & Cut C \ \ \ & Mass Cut \ \\ \hline
   $c\bar{b}\bar{c}b$ ($H^+H^-$+\,SM) & 0.99 \,fb & \ \,0.51 \,fb \\
   $c\bar{b} \bar{c}b$ (SM-only) &   0.29 \,fb & \ \,0.02 \,fb\\
   \hline
   $t\bar{t}$& 1.51 \,fb & \ \,0.03 \,fb \\
   $b\bar{b}jj$\,(QCD) & 0.77 \,fb & \ \,0.06 \,fb \\
   $c\bar{c}\tilde{j}\tilde{j}$\,(QCD) & 0.13 \,fb & \ \,0.02 \,fb \\
   {$b\bar{b}b\bar{b}$} & 0.27 \,fb & \ \,0.01 \,fb \\
   $ZZ$   & 0.15 \,fb & \ \,0.04 \,fb \\
   $Zh$   & 0.04 \,fb & \ \,0.01 \,fb  \\
   $W^+W^-$   & 0.55 \,fb & \ \,0.02 \,fb \\ 
   \hline
   Total (SM-only) & 3.72 \,fb & \ \,0.21 \,fb \\
   Total ($H^+H^-$+\,SM)  & 4.42 \,fb & \ \,0.68 \,fb \\
  %  $S/B$ & 0.1 & 0.16 \\
   Significance ($\mathcal{L} = 1\, \rm{ab}^{-1}$) \ \ \ & \ 11.1 & \ \ \ 26.3 \\
 \hline\hline
  \end{tabular}
  \caption{
  Same as Table ~\ref{tab:cutAB}, but after Cut C and Mass Cut.
  See text for more details.}
  \label{tab:cutCD}
\end{table}
%

%---------------------------------------------
%. Explanation for Sharp drop at MH
%----------------------------------------------

{\it \textbf{\boldmath ``Edge'' at $m_{H^+}$}.---} \ 
Cut C turns out to be very powerful at rejecting background
while suitably efficient in retaining our signal.
Inspecting the reconstructed $m_{bj}^{1,2}$ distributions reveal 
a peculiar ``Jacobian''-like drop of signal around 200~GeV, analogous to
what we see with $W \to \ell \nu$ transverse mass plot. 
Note that the cut of Eq.~(\ref{eq:pairing}) itself does not import any mass information.

One may notice that we have kept the $W^+W^- \to 4\tilde{j}$ 
background (magenta) and have not vetoed it.
This would practically be $c\bar s\bar cs$ with 
both charm mistagged as $b$-jets, as the ILD card of Delphes is able to do.
One can see the same behavior of an ``edge'' around the $W$ mass. 
This suggest that it is a common effect due to two-body production
in the C.M. frame, since at the ILC one collides $e^+$ and $e^-$ 
without parton distributions as at hadron colliders. 
As a result, we have exact knowledge of the kinematics, 
resulting in maximum possible available transverse momentum 
for the charged Higgs, approximated by 
$p_T^{{\rm max}}(H^+) = |\vec{p}_{H^+}| = \sqrt{s/4 - m^{2}_{H^{+}}}$, 
which is $\simeq 150\, (458)$\;GeV for 500\,GeV (1\,TeV) ILC. 
Beyond this there is barely any event, which creates a Jacobian-like fall 
after 200 (150)\;GeV for reconstructed $m_{bj}$ ($p_T^{H^+}$).

\begin{figure*}[t]
  \centering
  \includegraphics[scale=0.3]{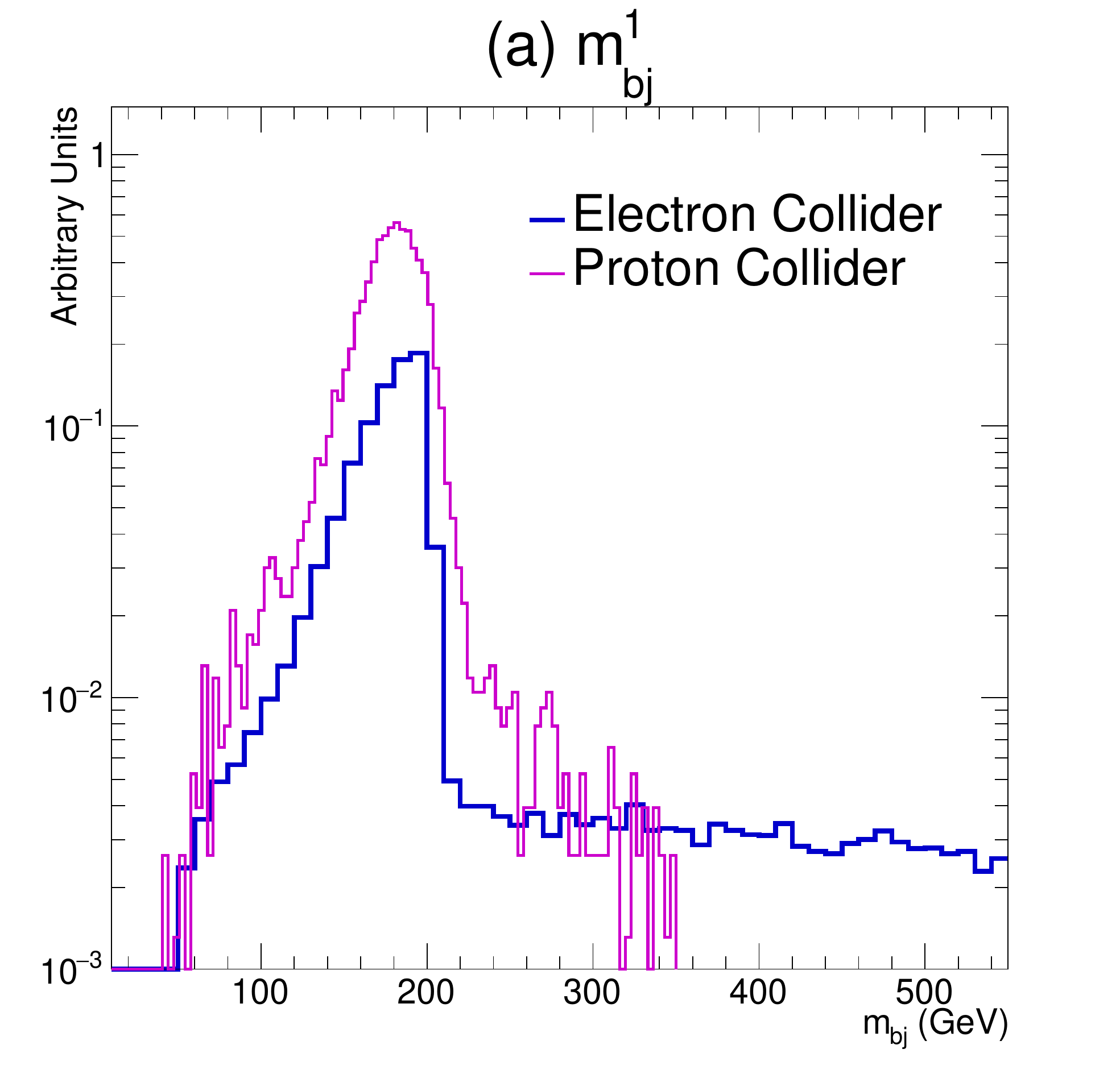}
  \includegraphics[scale=0.3]{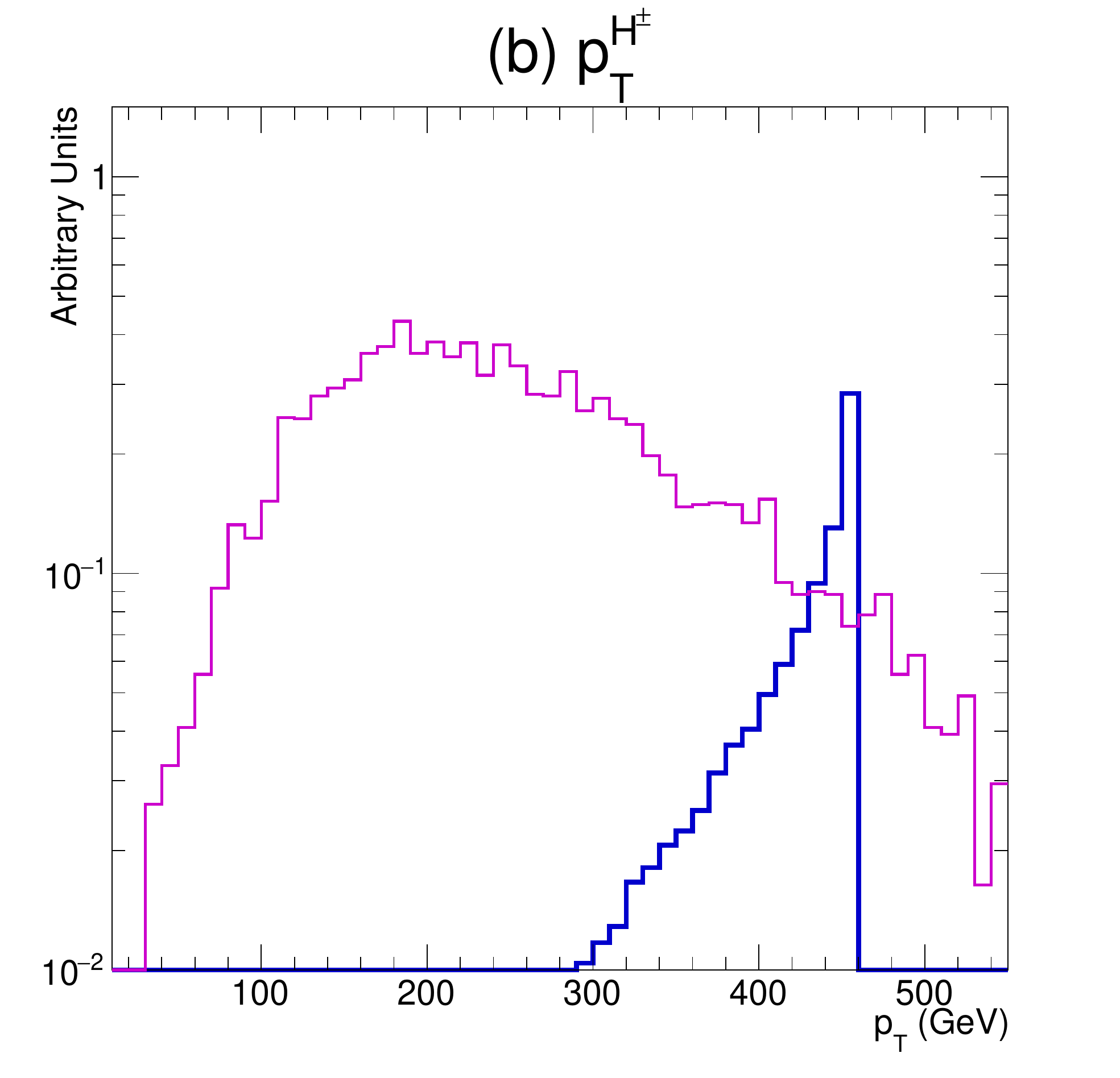}
  \caption{
Comparison of
 {(a)} reconstructed {\boldmath $m^{1}_{bj}$}, and
  (b) $p_T^{H^{\pm}}$ at {truth level} from
     $e^+ e^- \to H^+ H^- \to c\bar{b} \bar{c} b$ (blue) at $\sqrt{s} = 1$ TeV and
     $p p \to H^+ H^- \to c \bar{b} \bar{c} b$ (magenta) at $\sqrt{s} = 14$ TeV after Cuts A.
}
  \label{fig:eepp}
\end{figure*}

In case of $pp$ collider, %
{we expect some spread in distribution due to parton momenta.} 
To confirm this we reconstruct $m_{bj}$ using
the same procedure for a $pp$ collider with $\sqrt{s} = 14$\;TeV 
and compare with $e^+ e^-$ at $\sqrt{s} = 1$\;TeV. 
The results with arbitrary scale are compared in Fig.~\ref{fig:eepp}(a), 
along with a truth level $p_T^{H^+}$ from both $e^+ e^-$
and $pp$ colliders in Fig.~\ref{fig:eepp}(b). 
The $p_T^{H^+}$ at $e^+ e^-$ collider drops sharply at $\sim 460$\;GeV,
while for the $pp$ collider it is stretched out by parton distributions
as the $H^+H^-$ pair is not in the C.M. frame.
For the $m_{bj}$ distribution, we see the sharp ``edge'' at 200\;GeV
for $e^+e^-$ collider.
In comparison, for $pp$ collider, there is some hint of the ``edge''. 
However, given
{the leading order} cross section after Cuts\;A 
is only at 1.2 fb level at 14\;TeV, it would be completely 
swamped by much larger QCD background.

\begin{figure}[b]
  \centering
  \includegraphics[scale=0.3]{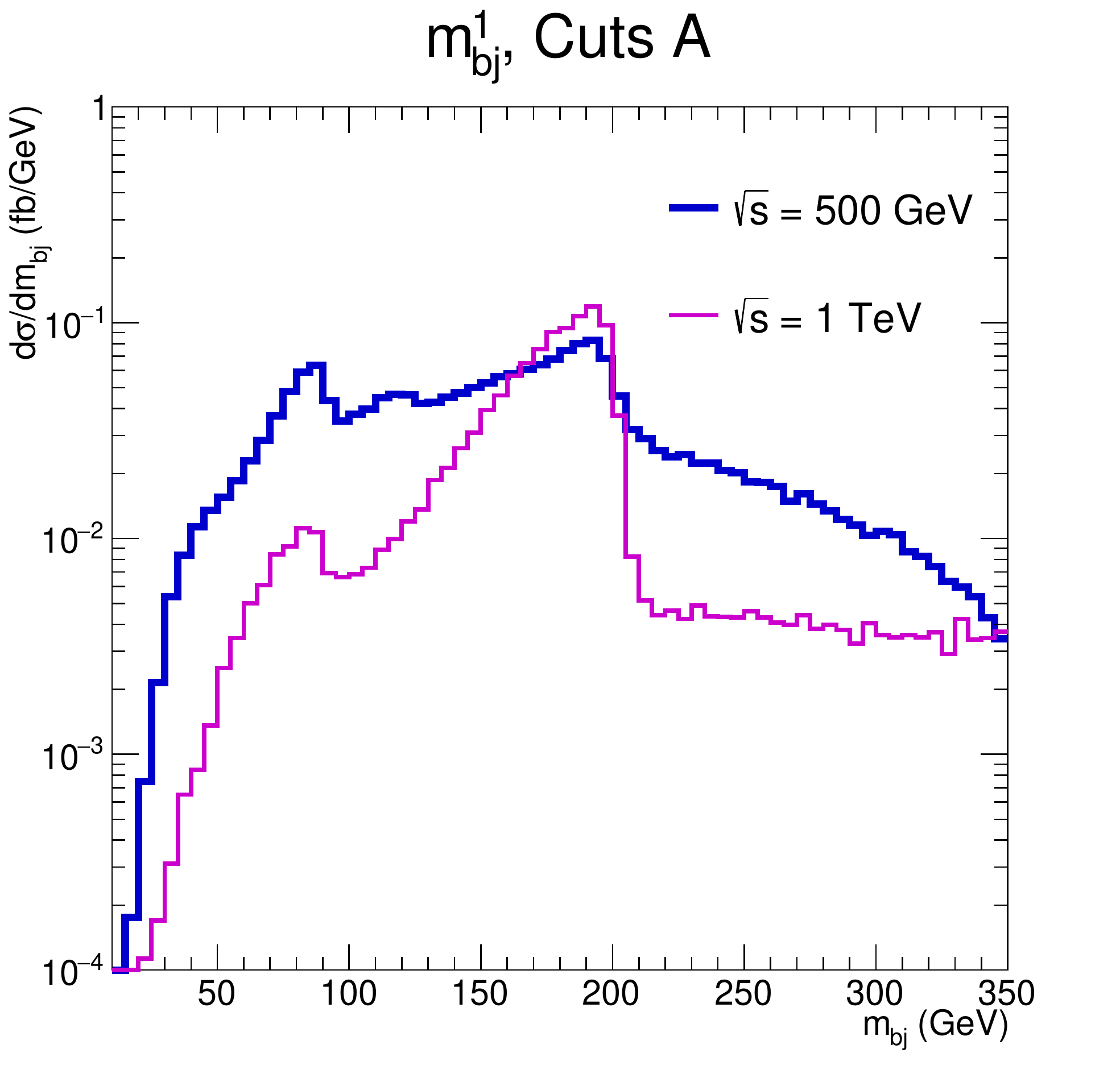}
\caption{
Reconstructed $m_{bj}^1$ after Cuts A for 500\;GeV (blue) 
and 1\;TeV (magenta) C.M. energies.}
  \label{fig:comp}
\end{figure}

{\it \textbf{Improvement with C.M. Energy}.---} \ 
The angular proximity requirement described as Cuts A would work 
better at higher C.M. energy, as the $H^+$ boson becomes more boosted. 
In Fig.~\ref{fig:comp} we show the reconstructed $m_{bj}^1$ 
after Cuts A for $\sqrt{s} = 500$\;GeV and 1\;TeV. 
We see clearly that the peak around $m_{H^{\pm}}$ 
becomes more prominent for 1\;TeV case than for 500\;GeV, 
%as our code identify more correct $bj$ pairs. 
and the ``edge'' also appears sharper. 
Further improvements can be achieved by applying a cut on the charged Higgs transverse momentum, but we do not explore this possibility further.

Following Fig.~\ref{fig:CutC}, one can extract $m_{H^+}$
from fitting ``the edge'', which we leave to future experiments to do.
We apply the mass window cut of 160\,GeV $ \leq m_{bj} \leq $ 210\,GeV 
on both pairs after Cut C, which we call the Mass Cut. 
The lower value is chosen by comparing with background,
which the experimental analysis can do much better.
We find that the significance jumps to 26.3 $\sigma$. 
The results after Cut C and the Mass Cut are presented in Table~\ref{tab:cutCD}.
While our result shows some scope for mass reconstruction but we have not 
included systematic uncertainties, uncertainties in the detector resolution etc. 
in our analysis.  We remark that such uncertainties would alleviate the possibility of $m_{H^+}$ reconstruction to some extent.
We leave out such a detailed analysis for future.  

%-----------------------------------------------------
%.  Figure 4, Comparison between ee and pp colliders
%-----------------------------------------------------
%-----------------

%--------------------------------------------
%. Figure 5, Mbj 0.5 TeV vs 1 TeV
%--------------------------------------------

\begin{figure*}
  \centering
   \includegraphics[scale=0.366]{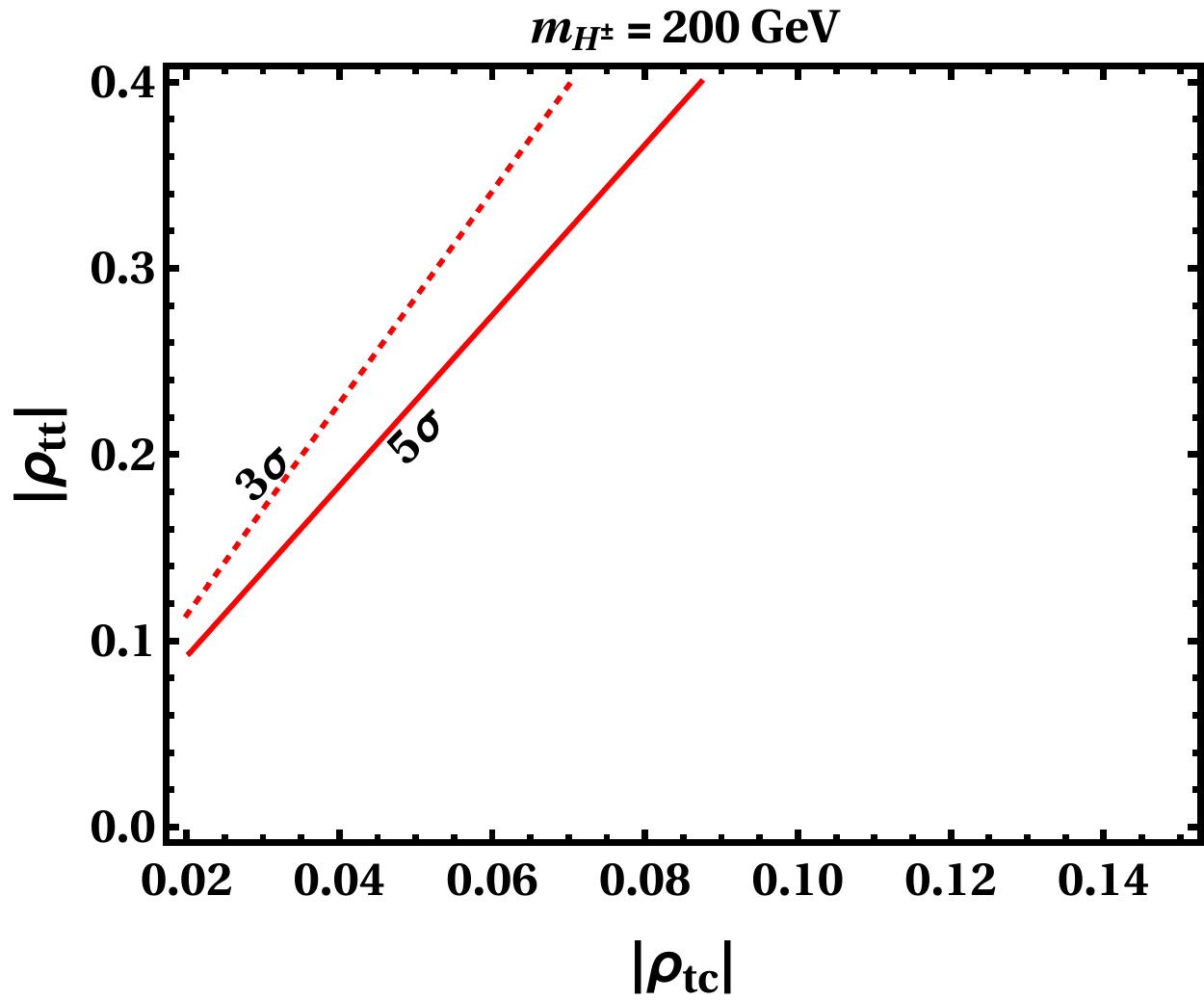} 
   \includegraphics[scale=0.366]{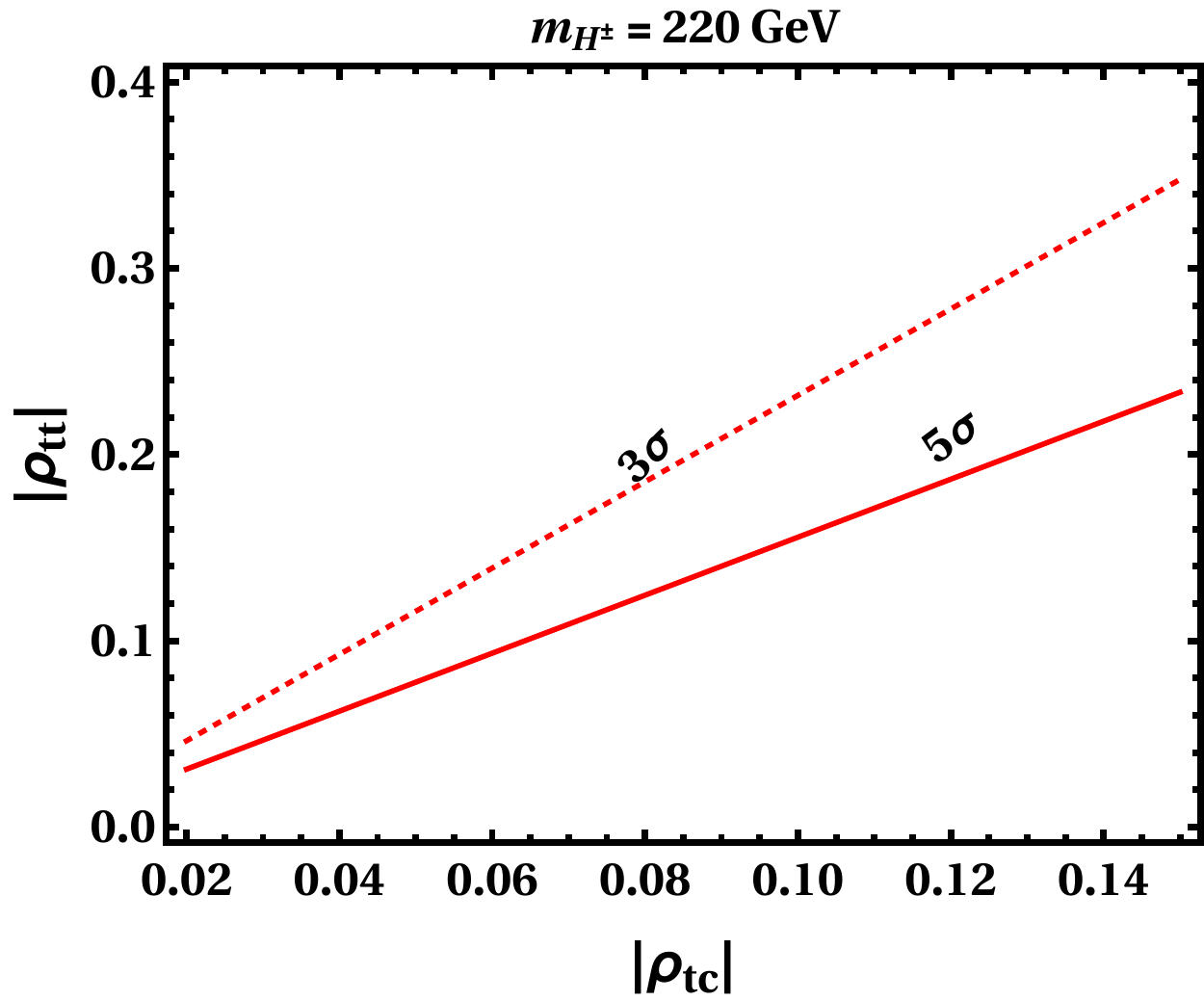}
   \includegraphics[scale=0.366]{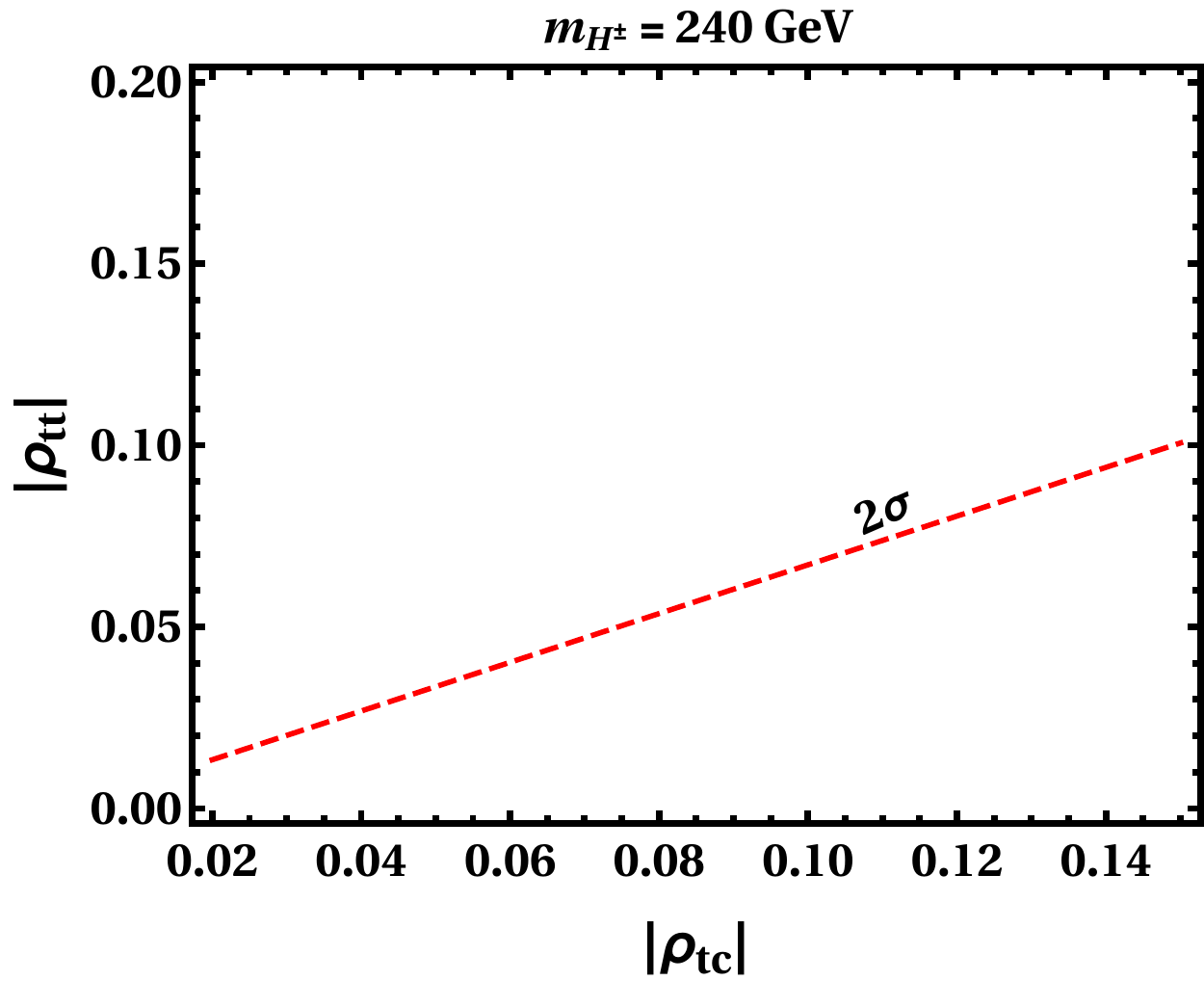}
\caption{The $5\sigma$ and $3\sigma$ contours in $\rho_{tt}$--$\rho_{tc}$ plane of
the $e^+e^- \to \gamma^*/Z^* \to H^+H^- \to c\bar{b}\bar{c}b$ process at $\sqrt{s}$ = 500 GeV for
$m_{H^+} = 200$ (left panel) and 220 GeV (middle panel) respectively. With the same CM energy 
we also plotted $2\sigma$ contour for $m_{H^+} = 240$ GeV in the right panel. Here we 
 assumed 1 ab$^{-1}$ integrated luminosity for $m_{H^+} = 200$ and 220 GeV but 10 ab$^{-1}$ for 
240 GeV. See text for details.}
 \label{fig:disccont}
\end{figure*}

Our focus of interest was a parameter space where all $\rho_{ij}$s are vanishingly small 
but $\rho_{tc}$ is nonzero with $m_{H^+} \simeq m_H \simeq m_A \simeq 200$ GeV. Such a parameter space renders a
blind spot for discovery at the LHC. We now discuss briefly the impact of different $\rho_{ij}$s 
on the discovery of Eq.~\eqref{eq:eeH+H-}. Presence of such couplings would alleviate the discovery potential
via suppression in the $H^+ /H^- \to c \bar b/ \bar c b$ branching ratios.
In general, $\rho_{ii}\sim \lambda_i$ i.e. $\rho_{tt} \sim \lambda_t$,
$\rho_{bb} \sim \lambda_b$ etc. As for example, it has been found that $|\rho_{tt}|\gtrsim 0.4$
is excluded for $m_{H^+} \simeq m_H \simeq m_A \simeq 200$ GeV via direct and indirect 
searches~\cite{Modak:2020fij,Lee:2021rzy}. Therefore the impact of $|\rho_{tt}|$ on discovery would 
be largest if we saturate the coupling to its upper limit 0.4.
As for illustration we plotted in the Fig.~\ref{fig:disccont} the $5\sigma$ 
and $3\sigma$ contours for the process in Eq.~\eqref{eq:eeH+H-}
at $\sqrt{s}= 500$ GeV for $m_{H^+} = 200, 220$ GeV and $2\sigma$ contour for 240 GeV in the $|\rho_{tt}|$--$|\rho_{tc}|$ plane.
While generating Fig.~\ref{fig:disccont} the selection cuts are kept as in before for $m_{H^+} = 200$ GeV but,
for $m_{H^+} =220$ (240) GeV we changed $180~\mathrm{GeV} < m_{bj} < 230$ GeV,
($200~\mathrm{GeV} < m_{bj} < 250$ GeV). In finding in Fig.~\ref{fig:disccont}
we rescaled the $\sigma\times \mathcal{B}(H^+/H^-\to c\bar b / \bar c b )$ of the process in Eq.~\eqref{eq:eeH+H-}
and ignored mild interference effects for simplicity. Further, for 
both $m_{H^+} = 200$ and 220 GeV we chose integrated luminosity 1 ab$^{-1}$. However, for 240 GeV due to the rapid fall in the cross section
near the threshold of the $\sqrt{s}= 500$ CM energy one may only have $2\sigma$ significance with 10 ab$^{-1}$ which is shown by the 
dashed red contour in the right panel of Fig.~\ref{fig:disccont}. It should also be noted that for a fixed value of $\rho_{tt}$ one may need a larger 
$\rho_{tc}$ for  $m_{H^+} = 220$ GeV compared $m_{H^+} = 200$ GeV. This is again simply due to the fall in cross sections as we approach to the $m_{H^+}=\sqrt{s}/2$ threshold.
It is clear from the Fig.~\ref{fig:disccont} that if $|\rho_{tt}|$ is large one would require larger $|\rho_{tc}|$ for discovery
for all three chosen masses. Note that achievable significance reduces when $|\rho_{tt}|$ becomes 
large for all three $m_{H^+}$ in Fig.~\ref{fig:disccont}. Impact of nonvanishing $\rho_{bb}$ is 
expected to be mild since it is more stringently constrained than $|\rho_{tt}|$ for our target mass ranges~\cite{Modak:2021vre}.
A discovery of  $e^+e^- \to \gamma^*/Z^* \to H^+H^- \to c\bar{b}\bar{c}b$
is possible for $m_{H^+}$ below our target mass range 200 GeV however for $m_{H^+}\ ,m_H\ , m_A < m_t$ we are subjected to additional 
constraints arising from different direct and indirect searches. 
E.g. if $m_A = m_H = m_{H^\pm} \lesssim m_t + m_c (m_b)$ GeV one may have processes such as $pp\to t \bar t$ followed
by $t \to c H/ A$, $t \to H^\pm b$ decays as potential discovery modes at the LHC. In addition, observables such as
$\mathcal{B}(B\to X_s \gamma)$ and $B_{s,d}$ mixing would provide stringent constraints for $\rho_{tc}$ via 
charged Higgs-top-charm-quark loop~\cite{Altunkaynak:2015twa}. In our exploratory analysis 
we have not considered such parameter space and leave out a detailed analysis for future.

%---------------------------------------------
%. Section IV
%----------------------------------------------

\section{Discussion and Conclusion}

We have given the scenario of relatively weak
$\rho_{tc}$ and $\rho_{tt}$ couplings 
--- still sufficient for EWBG for the latter ---
together with light $H$, $A$, $H^+$, degeneracy at $\sim 200$\:GeV,
that could evade LHC search.
That is, the $cg \to tH/tA \to tt\bar c$, $tt\bar t$ cross sections become too low,
while $cg \to bH^+$ production ends dominantly in $bc\bar b$ final state.
But the electroweak production of
$e^+e^- \to \gamma^*/Z^* \to H^+H^- \to c\bar b\bar cb$
not only can be discovered with high significance at the ILC operating at 500\;GeV,
but provide $H^+$ mass through this ``edge'', 
or sharp fall of $m_{bj}$ distribution after applying
the equal pair mass filter.
In the small $\rho_{tt}$ limit, the $c\bar b\bar cb$ electroweak production
through $e^+e^- \to H^+H^-$ is even independent of the $\rho_{tc}$ strength.
But in consideration of electroweak baryogenesis, 
we have kept $\rho_{tt} \sim {\cal O}(0.1)$,
which is still robust~\cite{Fuyuto:2017ewj} for generating BAU.
Such $\rho_{tt}$ strength can be measured via
the kinematically suppressed $t\bar b\bar cb$ production.

We have elucidated the cause for the ``edge'' that facilitates
the extraction of $m_{H^+}$ as due to the two-body kinematics
of $H^+H^-$ pair production in the center of mass frame.
For this purpose, we have not vetoed $W^+W^-$ production
in Figs.~\ref{fig:Kin} and \ref{fig:CutC}.
One clearly sees from Figs.~\ref{fig:CutC} the ``edge'' at $M_W$
from the $c\bar s\bar cs$ final state, which can be used as a demonstration mode. 
Furthermore, $W^+W^- \to c\bar s \ell\nu$ can be fully reconstructed,
and one can clearly separate the two $W$'s as further check.

The latter point brings about a comment of curiosity.
The muon $g-2$ anomaly has recently been confirmed~\cite{Muong-2:2021ojo} 
by the muon g-2 collaboration at FNAL.
It has been suggested~\cite{Hou:2021sfl} that it could be 
explained by the one-loop mechanism in g2HDM 
with sizable $\rho_{\tau\mu} \simeq \rho_{\mu\tau}$,
where for $m_{H,A} \sim 200$\;GeV,
the strength does not need to be as large as 0.2.
On one hand this illustrates the versatility and richness
of the phenomenology of g2HDM, while
on the other hand one might have the $c\bar b\tau\nu$ discovery mode 
analogous to the $c\bar s\tau\nu$ for $W^+W^-$ that we just mentioned.
However, the $\rho_{\tau\mu} \simeq \rho_{\mu\tau}$ coupling strength 
would be larger than our $\rho_{tc},\, \rho_{tt}$ values, 
hence the $H,\, A,\, H^+$ bosons would likely be discovered 
already at the LHC, as discussed in Ref.~\cite{Hou:2021sfl},
and the ILC would be needed less.

We can also have flavor diagonal process like $e^+ e^- \to Z^* \to H A \to 4b, 4\tau, 2b2\tau$,
but these process require a non-zero $\rho_{bb}$ and $\rho_{\tau\tau}$. 
In this study we have $\rho_{bb},\  \rho_{\tau\tau} \to 0$ hence these channels
are highly suppressed. Interested reader can look at Ref~\cite{Modak:2021vre} where they have 
considered $4b$ final state at ILC, in a $\rho_{bb}$ driven EWBG scenario.

In conclusion, if extra top Yukawa couplings turn out weak
while the extra $H, A, H^+$ bosons turn out light, even the HL-LHC 
may not be able to uncover them.
We show, with example of $m_H \sim m_A \sim m_{H^+} \sim 200$\;GeV,
that the electroweak production of $e^+e^- \to H^+H^- \to c\bar b\bar cb$
may not only reveal the presence of such physics,
but allow the extraction of $m_{H^+}$.
The $t\bar b\bar cb$ mode can provide information on $\rho_{tt}$
and check whether electroweak baryogenesis is achievable.
The presence of $H$ and $A$ can be revealed in same-sign top, 
or $tt\bar c$ ($\bar t\bar t c$) production,
but how to access $m_H$, $m_A$ needs further study.
\\ \\
\noindent{\bf Acknowledgments}\;
WSH and RJ are supported by MOST 110-2639-M-002-002-ASP of Taiwan, 
with WSH in addition by NTU 110L104019 and 110L892101.
TM is supported by a postdoctoral research fellowship 
from the Alexander von Humboldt Foundation.

%\clearpage

%---------------------------------------------------
% References
%---------------------------------------------------

%\end{thebibliography}
\end{document}